\documentclass[aps,amsmath,notitlepage,twocolumn,amssymb,prl,longbibliography]{revtex4-1}
\usepackage{graphicx}
\usepackage{dcolumn}
\usepackage{bm}
\usepackage[colorlinks=true,linkcolor = blue,citecolor=blue,urlcolor=blue,anchorcolor = blue]{hyperref}
\usepackage{wasysym}
\usepackage{stmaryrd}
\usepackage{verbatim}
\usepackage{subfigure}
\usepackage{amsmath}
\usepackage{times}
\usepackage[version=4]{mhchem}
\usepackage{braket}
\usepackage{siunitx}
\usepackage{amssymb}
\usepackage{color}
\usepackage{multirow}
\usepackage{tabularx}
\usepackage{amsthm,amsmath,amssymb}
\usepackage{mathrsfs}

\newcommand{\RNum}[1]{\uppercase\expandafter{\romannumeral #1\relax}}

\newcolumntype{Y}{>{\centering\arraybackslash\hsize=.6\hsize}X}
\newcolumntype{Z}{>{\centering\arraybackslash\hsize=.13\hsize}X}

\begin{document}
\title{Anomalous enhancement of magnetism by non-magnetic doping in the honeycomb-lattice antiferromagnet \ce{ErOCl}}

\author{Yanzhen\,Cai$^{1,2}$}
\author{Mingtai\,Xie$^{1,2}$}
\author{Jing\,Kang$^{2}$}
\author{Weizhen\,Zhuo$^{1,2}$}
\author{Wei\,Ren$^{1,2}$}
\author{Xijing\,Dai$^{1}$}
\author{Anmin\,Zhang$^{1}$}
\author{Jianting\,Ji$^{2}$}
\author{Feng\,Jin$^{2}$}
\author{Zheng\,Zhang$^{2}$}
\email{zhangzheng@iphy.ac.cn}
\author{Qingming\,Zhang$^{1,2}$}
\email{qmzhang@iphy.ac.cn}

\affiliation{$^{1}$School of Physical Science and Technology, Lanzhou University, Lanzhou 730000, China}
\affiliation{$^{2}$Beijing National Laboratory for Condensed Matter Physics, Institute of Physics, Chinese Academy of Sciences, Beijing 100190, China}

\begin{abstract}
Tuning magnetic anisotropy through chemical doping is a powerful strategy for designing functional materials with enhanced magnetic properties. 
Here, we report an enhanced Er$^{3+}$ magnetic moment resulting from non-magnetic Lu$^{3+}$ substitution in the honeycomb-lattice antiferromagnet ErOCl.
Unlike the Curie-Weiss type divergence typically observed in diluted magnetic systems, our findings reveal a distinct enhancement of magnetization per Er$^{3+}$ ion under high magnetic fields, suggesting an unconventional mechanism.
Structural analysis reveals that Lu$^{3+}$ doping leads to a pronounced contraction of the $c$-axis, which is attributed to chemical pressure effects, while preserving the layered SmSI-type crystal structure with space group $R\overline{3}m$.
High-resolution Raman spectroscopy reveals a systematic blueshift of the first and seventh crystalline electric field (CEF) excitations, indicating an increase in the axial CEF parameter $B_{2}^{0}$.
This modification enhances the magnetic anisotropy along the $c$-axis, leading to a significant increase in magnetization at low temperatures and under high magnetic fields, contrary to conventional expectations for magnetic dilution.
Our work not only clarifies the intimate connection between magnetism and CEF in rare-earth compounds, but more importantly, it reveals a physical pathway to effectively tune magnetic anisotropy via anisotropic lattice distortion induced by chemical pressure.
\end{abstract}

\maketitle

\subsection{Introduction}
Controlling and tuning magnetic anisotropy is a central theme in modern condensed matter physics, as it governs the stability of magnetic states and the functionality of magnetic materials~\cite{1,2,3}. 
In this endeavor, rare-earth-based compounds have emerged as a particularly rich platform, owing to their strong spin-orbit coupling (SOC) and a rich spectrum of crystalline electric field (CEF) states~\cite{7,8,9}.
These characteristics endow rare-earth magnets with strong
magnetic anisotropy, offering a powerful avenue for tailoring
their properties via external perturbations, including hydrostatic
pressure, strain, and chemical doping~\cite{4,5,6,39,40}.

Among rare-earth compounds, layered materials have emerged as a focal point, encompassing rare-earth chalcogenides AReCh2 (A = alkali or monovalent ions, Re = rare earth, Ch = O, S, Se)~\cite{10,11}, chalcohalides ReChX (Re = rare earth, Ch = O, S, Se, Te, X = F, Cl, Br, I)~\cite{12}, and chlorides YbCl$_{3}$~\cite{13,14}.
Their quasi-two-dimensional crystal structures render them highly sensitive to perturbations, which modify interlayer spacing or coordination environments, consequently affecting exchange interactions and CEF excitations~\cite{41,42}.
For example, in Ce-based magnetic materials, altering coordinating anions modulates spin exchange interactions, leading to modifications in ground-state magnetism and low-energy spin excitations~\cite{15}.
Additionally, dilute magnetic doping, which substitutes magnetic ions with non-magnetic ones, is a common strategy for tuning magnetic properties.
However, this process typically disrupts magnetic ordering and reduces magnetization by introducing disorder into the magnetic system~\cite{16}.
Nevertheless, recent studies on low-dimensional and geometrically frustrated systems reveal that non-magnetic doping can induce novel effects, such as unconventional spin textures in triangular lattices~\cite{17}.
Moreover, doping offers valuable insights into the evolution of ground-state magnetism, shedding light on the mechanisms governing complex magnetic states~\cite{18}.
Therefore, dilute magnetic doping provides an effective means to probe the evolution of ground-state magnetism, yielding valuable insights into the mechanisms governing complex magnetic states.

In recent years, ReOCl (Re = rare earth) compounds have attracted increasing attention~\cite{47}. Among them, ErOCl shows great potential for applications in optoelectronic modulation, information encryption, thermal imaging, and high-performance device integration~\cite{48,49,50}. However, current studies remain at a preliminary stage, and its magnetic properties have yet to be thoroughly explored. As a rare-earth compound, ErOCl is expected to exhibit rich magnetism governed by CEF. Therefore, we performed a detailed investigation of its magnetic behavior.

To determine the complete CEF scheme underlying magnetic anisotropy, Raman spectroscopy provides a powerful and efficient approach. Compared with inelastic neutron scattering (INS), it offers higher energy resolution, requires only small sample volumes, and can be implemented with standard laboratory instrumentation. Raman spectroscopy directly probes electronic transitions between CEF levels, with selection rules dictated by wavefunction symmetry. Polarization-resolved measurements enable these rules to be tested and the symmetry of each CEF state to be identified. This methodology, well established in seminal studies of rare-earth phosphates~\cite{34}, forms the basis of our investigation of ErOCl.

In this work, we successfully synthesized high-quality bulk ErOCl single crystals with dimensions of $10$ mm, providing a solid foundation for studying its magnetic properties. ErOCl crystallizes in the YbOCl/SmSI-type structure~\cite{12,19,20,21} (space group $R\overline{3}m$), where Er$^{3+}$ ions form a honeycomb lattice stacked along the $c$-axis with weak interlayer interactions. This structural feature enables ErOCl to be regarded as a well-defined van der Waals magnetic material.

Such a configuration naturally results in strong magnetic anisotropy governed by the CEF environment around the {Er$^{3+}$ sites.
	Here, we reveal that substituting Er$^{3+}$ with non-magnetic Lu$^{3+}$ unexpectedly enhances magnetization at low temperatures, in stark contrast to conventional magnetic dilution effects.
	By systematically synthesizing a series of Lu$_{x}$Er$_{1-x}$OCl} single crystals and performing complementary structural, spectroscopic, and magnetization measurements, we demonstrate how chemical pressure—primarily resulting from the reduction of the $c$-axis lattice constant—amplifies the key uniaxial CEF parameter $B_{2}^{0}$.
This enhancement of the uniaxial magnetic anisotropy ultimately leads to the observed strengthening of magnetization under an external magnetic field.

To elucidate the microscopic mechanisms underlying this phenomenon, we employ high-resolution X-ray diffraction (XRD) and Raman scattering.
XRD confirms that all Lu-substituted samples remain isostructural with ErOCl but exhibit distinct lattice shrinkage along the $c$-axis.
Our Raman measurements further confirm that the first and seventh CEF excited levels shift to higher energy with increasing Lu$^{3+}$ content, indicating a modification of the CEF environment.
A detailed analysis of the Raman and magnetization data identifies $B_{2}^{0}$ as the most responsive parameter to $c$-axis lattice compression.
Concurrently, magnetization measurements reveal that the low temperature magnetization—particularly along the $c$-axis under strong magnetic fields—systematically increases with Lu$^{3+}$ content.
Such a correlated evolution of structural, spectroscopic, and magnetic properties highlights an effective route to deliberately tune CEF-driven anisotropy via non-magnetic doping.

Our work demonstrates that leveraging chemical pressure to tailor uniaxial CEF parameters is an effective method to achieve precise control over magnetic anisotropy, offering crucial physical insights for the rational design of rare-earth-based quantum materials. 
In the following section, we will conduct a detailed study on the structure, spectra and magnetism of ErOCl and Lu$_{x}$Er$_{1-x}$OCl and discuss the significance of our research results in the context of advanced rare-earth magnetic materials.

\begin{figure*}[t!]
	\centering
	\includegraphics[scale=0.95]{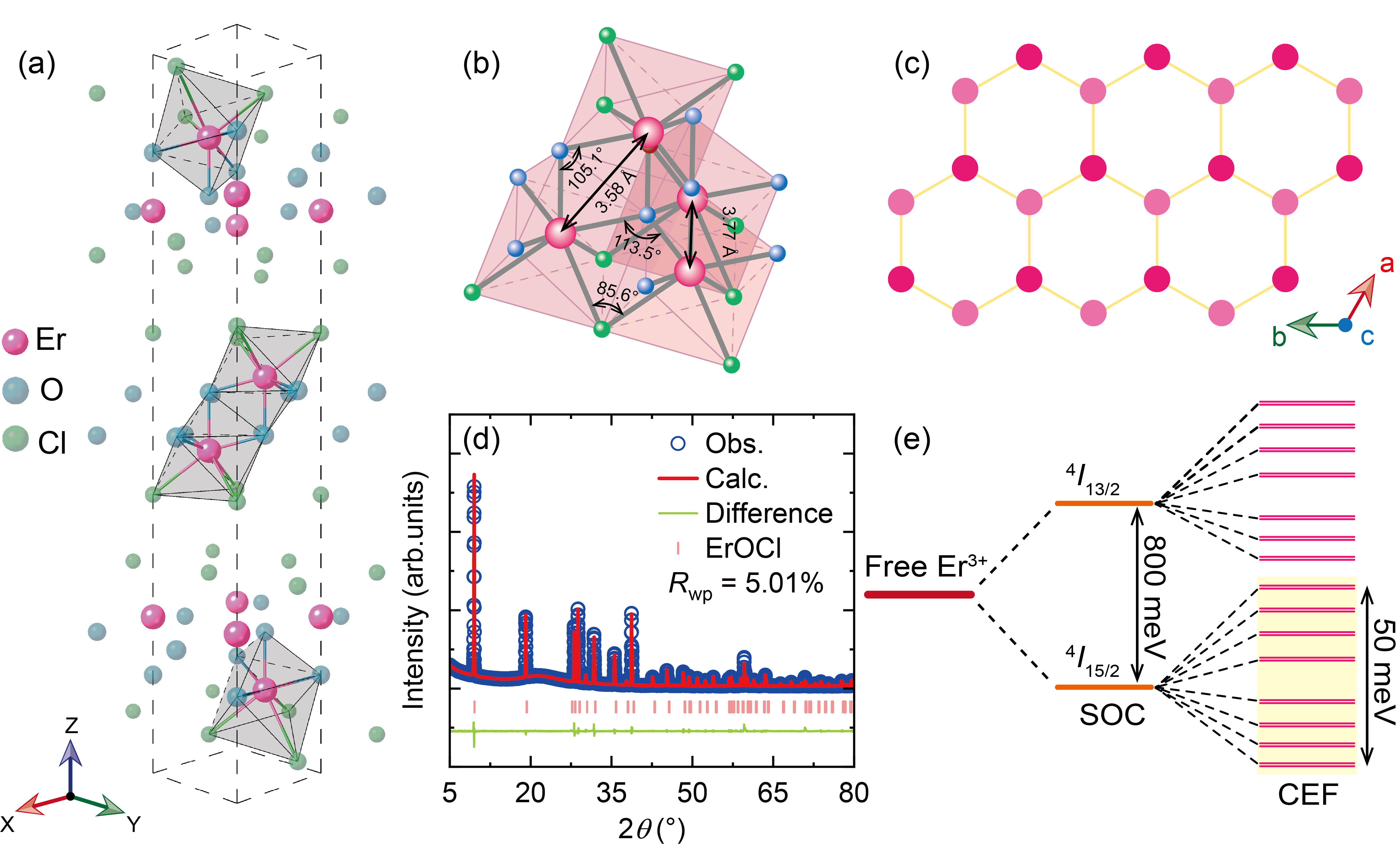}
	\caption{\textbf{Crystal structure of ErOCl, polycrystalline XRD pattern, and schematic diagram of CEF energy levels.} 
		(a) Crystal structure of ErOCl (space group $R\overline{3}m$), showing the layered structure formed by ErO$_4$Cl$_3$ polyhedra. 
		(b) The sevenfold coordination of Er$^3+$ by four O$^2-$ and three Cl$^-$. 
		(c) The bond lengths and bond angles of both NN and NNN magnetic ions are labeled.
		(d) Powder XRD pattern at 300 K showing observed data (blue open circles), Rietveld refinement result (red solid line), and difference (green line). 
		Light red tick marks indicate expected Bragg peak position. 
		The quality of the Rietveld refinement is indicated by $R_{wp} = 5.01\%$. 
		(e) Free Er$^{3+}$ ions first experience splitting of the spectral terms due to SOC, followed by further splitting of the CEF levels in the CEF environment. 
		For ErOCl, the energy span between the CEF ground state and the highest excited state of the $^{4}I_{15/2}$ term is approximately 50 meV.}
	\label{fig:Fig1}
\end{figure*}

\section{Samples, experiments, and methods}
Crystal Growth and Preparation: Single crystals of ErOCl were grown using the anhydrous ErCl$_3$ flux method~\cite{52}. 
The typical size of these single-crystal samples is 5 $\times$ 5 $\times$ 0.05 mm$^{3}$, with a maximum size of 10 mm.
The synthesis procedure for single-crystal samples of LuOCl and Lu$_{x}$Er$_{1-x}$OCl ($0 < x < 1$) follows a similar process. The synthesized samples were treated with deionized water to remove surface impurities, ultimately yielding transparent single crystals.\\

Sample characterization: The crystal structure of ErOCl, LuOCl, and Lu$_{x}$Er$_{1-x}$OCl ($0 < x < 1$) was determined through Rietveld refinement using the GSAS-II program from XRD patterns measured at 300 K~\cite{43}.
The data were collected using a Rigaku SmartLab X-ray diffractometer.
The detection range was 5°–80°, with a step size of 0.01°.
The crystal structure of these single-crystal samples was re-examined using a Rigaku XtaLAB Synergy four-circle diffractometer.
Scanning electron microscopy (SEM) was performed using a Hitachi Schottky Field Emission Scanning Electron Microscope SU5000 equipped with a Bruker XFlash 6-60 energy-dispersive spectrometer to determine the elemental ratios of ErOCl, LuOCl, and Lu$_{x}$Er$_{1-x}$OCl ($0 < x < 1$) single crystals.\\

Thermodynamic characterization: Approximately 1.2 mg of ErOCl and Lu$_{x}$Er$_{1-x}$OCl ($0 < x < 1$) single-crystal samples were prepared for bulk magnetization measurements, including temperature-dependent magnetization ($M/H$–$T$) and magnetic field-dependent magnetization ($M$–$H$) measurements.
These measurements were performed using a Quantum Design (QD) Physical Property Measurement System (PPMS) under magnetic fields from 0 to 14 T.
Anisotropic measurements were conducted along the $c$-axis and within the $ab$-plane over a temperature range of 1.8–100 K.\\

Raman spectroscopy: The Raman spectra were collected using an HR800 Evolution (Jobin Yvon) system equipped with a 473 nm laser, a charge-coupled device (CCD), and volume Bragg gratings.
After cleavage, single-crystal samples of ErOCl, LuOCl, and Lu$_{x}$Er$_{1-x}$OCl ($0 < x < 1$) were placed in a closed-cycle cryostat (AttoDRY 2100) equipped with a superconducting magnet capable of generating up to 9 T.
The excitation laser beam was focused to a spot of $\sim$5 $\mu$m in diameter on the $ab$-plane of single-crystal samples.\\

\subsection{Crystal structure of \ce{ErOCl}}
The rare-earth chalcohalide ErOCl adopts the SmSI-type structure within the trigonal system (space group $R\overline{3}m$), as illustrated in Fig. \hyperref[fig:Fig1]{1(a)}. The crystal structure of ErOCl was determined from powder XRD data via Rietveld refinement, as shown in Fig. \hyperref[fig:Fig1]{1(d)}. In addition, XRD measurements performed on as-grown ErOCl crystals reveal only $(00l)$ reflections [see Fig. \hyperref[fig:Fig4]{4(a)}], further confirming their high crystallinity and preferred orientation along the $c$-axis. The refined lattice parameters were determined to be $a = b = 3.769$ \AA~and $c = 27.931$ \AA, in agreement with the anticipated crystallographic model.

Two adjacent triangular magnetic layers are a structural feature of ErOCl, similar to those found in YbOCl. The nearest-neighbor (NN) Er$^{3+}$ magnetic ions are spaced 3.58 \AA~apart, closely resembling the next-nearest-neighbor (NNN) distance of 3.77 \AA, as illustrated in Fig. \hyperref[fig:Fig1]{1(b)}. When viewed along the $c$-axis, these magnetic ions form a compact honeycomb lattice structure, as depicted in Fig. \hyperref[fig:Fig1]{1(c)}. Within this unique lattice environment, the rich landscape of CEF excitations presents a compelling avenue of study. The well-defined energy levels of these CEF states make them highly accessible for external tuning, providing a direct physical pathway to precisely manipulate the material's macroscopic magnetic properties.

\begin{figure*}[t!]
	\includegraphics[scale=0.6]{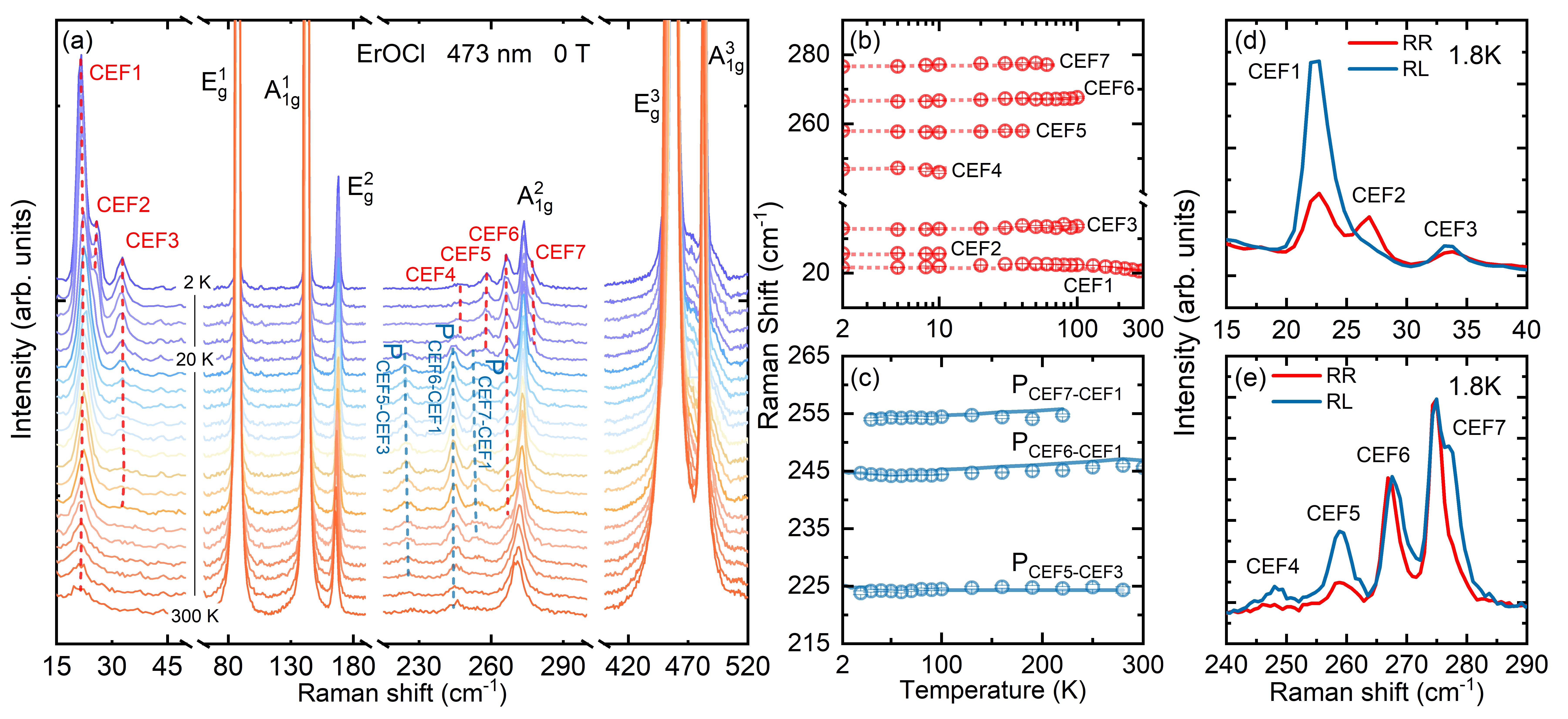}
	\caption{\label{fig:epsart}{ \textbf{Raman scattering spectra and CEF excitations in ErOCl}. 
			(a) The Raman scattering spectra of ErOCl at different temperatures.  Six Raman-active phonon modes are clearly observed and are labeled in black near the corresponding peaks.
			Seven CEF excitation peaks from the ground state to various excited states are marked with red dashed lines.
			Three excitation peaks between CEF excited states are observed and marked with blue dashed lines.
			(b) Temperature dependence of CEF excitations from the ground state to various excited states. For the Stokes process, Raman intensity increases as the ground state occupation number rises with decreasing temperature, consistent with the enhancement seen in (a) \cite{52}.
			(c) Temperature dependence of excitations between CEF excited states. The blue open circles are experimental data. The blue solid lines represent the calculated results based on the CEF eigen excitations in (b). The disappearance of those peaks at low temperatures in (a) corresponds to a decrease in the occupation number of excited states.
			(d)-(e) The Raman scattering spectra of ErOCl at 1.8 K. RL and RR represent cross-circular and parallel-circular polarization configurations, respectively.
	}}
	\label{fig:Fig2}
\end{figure*}

\subsection{CEF excitations in ErOCl}
 The electron configuration of the 4$f$ orbital in Er$^{3+}$ is 4$f^{11}$. SOC splits this configuration into multiple states, with $^{4}I_{15/2}$ as the ground state and $^{4}I_{13/2}$ as the first excited state. The energy gap between these multiplets is approximately 800 meV ($\sim$9000 K)~\cite{25}, ensuring that at low temperatures, the magnetic properties are primarily governed by the ground-state multiplet $^4I_{15/2}$~\cite{44,45}. 
Under the influence of the CEF and time-reversal symmetry, this multiplet further splits into a series of Kramers doublets [see Fig. \hyperref[fig:Fig1]{1(e)}], whose energy distribution dictates both the optical and magnetic behavior of Er$^{3+}$. In ErOCl, the Er$^{3+}$ ion is coordinated by three Cl$^{-}$ and four O$^{2-}$ anions, forming a local CEF environment with $C_{3v}$ point group symmetry, as depicted in Figs. \hyperref[fig:Fig1]{1(a)} and \hyperref[fig:Fig1]{1(b)}. Consequently, the CEF Hamiltonian that describes the excitations in ErOCl is expressed as~\cite{19}:
\begin{equation}
	\hat{H}_{CEF} =  \sum_{i} B_{2}^{0} \hat{O}_{2}^{0} + B_{4}^{0} \hat{O}_{4}^{0} + B_{4}^{3} \hat{O}_{4}^{3} + B_{6}^{0} \hat{O}_{6}^{0} + B_{6}^{3} \hat{O}_{6}^{3} + B_{6}^{6} \hat{O}_{6}^{6}, 
	\label{for:1}
\end{equation}
where $B_m^n$ are the CEF parameters, and $\hat{O}_m^n$ represents the Stevens operators constructed from the total angular momentum $\hat{J}$ following SOC. This formalism provides a comprehensive framework for understanding the electronic structure and magnetic behavior of Er$^{3+}$ in ErOCl.

The CEF excitations in ErOCl can be effectively probed through Raman scattering. Fig. \hyperref[fig:Fig2]{2(a)} presents the temperature-dependent Raman spectra of ErOCl recorded from 2 K to 300 K. At 2 K, six well-defined phonon peaks are observed at 87.7 (\( E_g^1 \)), 143.1 (\( A_{1g}^1 \)), 167.8 (\( E_g^2 \)), 273.5 (\( A_{1g}^2 \)), 458.5 (\( E_g^3 \)), and 483.9~cm$^{-1}$ (\( A_{1g}^3 \)).
Polarization-dependent Raman measurements further confirm the symmetry assignments of these phonon modes~\cite{52}. Notably, similar phonon features have been widely reported in the isostructural compounds YbOCl~\cite{21, 27} and LuOCl~\cite{28}, reinforcing the structural consistency across this family of materials. Beyond these phonon excitations, an intriguing observation is the emergence of additional peaks in the Raman spectra at lower temperatures. These extra peaks become increasingly pronounced as the temperature decreases, suggesting their origin beyond conventional lattice vibrations. Based on the following arguments, we attribute these additional features to CEF excitations in ErOCl, revealing the rich low-energy excitations of this material.

(i) ErOCl is a well-established insulator, eliminating the possibility of electronic continuum contributions. Within the energy range spanning from tens to hundreds wavenumbers, any observed excitation peaks must originate from lattice vibrations, magnetic excitations, or CEF excitations. By systematically identifying the phonon peaks, we ruled out the possibility that these additional excitations arise from lattice vibrations. Furthermore, no structural or magnetic phase transition was detected down to 1.8 K, excluding the likelihood of these excitations being associated with collective magnetic order. CEF excitations of Er$^{3+}$ ions typically fall within a few to a few tens of meV, as reported in compounds such as KErSe$_2$~\cite{29}, CsErSe$_2$~\cite{29}, and NaErS$_2$~\cite{30}. Given this energy scale, the additional peaks observed in the Raman spectra align well with CEF excitations, further supporting their origin as electronic transitions within the CEF split states of Er$^{3+}$.

(ii) By analyzing the temperature-dependent Raman spectra, we identified a total of ten additional excitation peaks, whose evolution with temperature is highlighted by red and blue dashed lines in Fig. \hyperref[fig:Fig2]{2(a)}. These peaks can be categorized into two distinct types. The first type, highlighted in red, corresponds to intrinsic CEF excitations, representing transitions from the CEF ground state to various excited states. The second type, depicted in blue, arises from transitions between thermally populated CEF excited states, classified as non-intrinsic excitations. These two types of excitations exhibit distinct characteristics. Intrinsic CEF excitations remain clearly visible even at the lowest measured temperature of 2 K. In contrast, non-intrinsic excitations are nearly undetectable below 20 K, as shown in Figs. \hyperref[fig:Fig2]{2(b)} and \hyperref[fig:Fig2]{2(c)}. The temperature dependence of these non-intrinsic excitations can be quantitatively determined by subtracting lower-energy CEF excitation levels from higher-energy ones at the same temperature. The three solid lines in Fig. \hyperref[fig:Fig2]{2(c)} illustrate this approach, where the energy differences between the seventh and first, sixth and first, and fifth and third CEF excitation states successfully reproduce the temperature-dependent evolution of the corresponding non-intrinsic excitations. Furthermore, we found that both types of CEF excitations exhibit negligible shifts with temperature, reinforcing their identification as CEF excitations in ErOCl. This clear distinction between intrinsic and non-intrinsic excitations, as well as their consistent temperature evolution, provides strong experimental validation for the CEF scheme in this system.

(iii) Leveraging the CEF energy levels identified from Raman scattering and the CEF Hamiltonian described in Eq. \hyperref[for:1]{(1)}, we determined the optimal CEF parameters for ErOCl. Using the CEF parameters predicted by the point charge model as initial values, we refined the CEF parameters through least-squares fitting, as summarized in Table \ref{table:TABLE I}. By diagonalizing the CEF Hamiltonian with these optimized parameters, we obtained the CEF energy levels and corresponding wave functions of ErOCl, as detailed in Table~\ref{table:TABLE II}. The excellent agreement between the fitted CEF energy levels and those observed in Raman scattering provides a robust, self-consistent validation that the additional excitation peaks in the Raman spectra indeed originate from CEF excitations in ErOCl.

(iv) Beyond determining the CEF energy levels, a deeper analysis of the CEF wave functions in ErOCl can be conducted through symmetry considerations. In the $C_{3v}$ CEF environment, the eight doubly degenerate CEF states of Er$^{3+}$ split into distinct irreducible representations, classified as $5\Gamma_4 + 3\Gamma_{5,6}$. From the diagonalized CEF wave functions, we identified that the ground state, second, third, fifth, and sixth excited states belong to the $\Gamma_4$ representation, whereas the first, fourth, and seventh excited states belong to $\Gamma_{5,6}$. This symmetry classification follows the Raman transition selection rules~\cite{31, 32}:

\begin{equation}
	\Gamma_i \otimes \Gamma_f \subseteq \Gamma_{Raman},
	\label{for:2}
\end{equation}

{where \( \Gamma_i \) and \( \Gamma_f \) denote the symmetries of the initial and final states, and  \( \Gamma_{Raman} \) represents the symmetry of the Raman scattering tensor. The allowable transitions for ErOCl can be decomposed as follows~\cite{33, 34}:
	
	\begin{equation}
		\Gamma_4 \to \Gamma_4 = \Gamma_1 \oplus \Gamma_2 \oplus \Gamma_3,
		\label{for:3}
	\end{equation}
	
	\begin{equation}
		\Gamma_4 \to \Gamma_{5,6} = \Gamma_3,
		\label{for:4}
	\end{equation}
	
	The corresponding Raman tensors for $\Gamma_1$, $\Gamma_2$, and $\Gamma_3$ are provided in the Supplementary Material \cite{52}. Using these tensors, the expected Raman scattering intensities for cross-circular (RL) and parallel-circular (RR) polarizations can be derived. For transitions within the $\Gamma_4$ manifold ($\Gamma_{4}$ $\rightarrow$ $\Gamma_{4}$),  the Raman scattering intensity follows:
	
	\begin{equation}
		I^{RR}_{\Gamma_4 \rightarrow \Gamma_4} = a^2 + c^2,
		\label{for:5}
	\end{equation}
	
	\begin{equation}
		I^{RL}_{\Gamma_4 \rightarrow \Gamma_4} = 8d^2,
		\label{for:6}
	\end{equation}
	
	In contrast, for transitions between different irreducible representations ($\Gamma_{4}$ $\rightarrow$ $\Gamma_{5,6}$), the selection rules dictate:
	\begin{equation}
		I^{RR}_{\Gamma_4 \rightarrow \Gamma_{5,6}} = 0,
		\label{for:7}
	\end{equation}
	
	\begin{equation}
		I^{RL}_{\Gamma_4 \rightarrow \Gamma_{5,6}} = 8d^2,
		\label{for:8}
	\end{equation}
	
	These selection rules provide a rigorous framework to interpret the polarization-resolved Raman spectra of ErOCl. In Figs. \hyperref[fig:Fig2]{2(d)} and \hyperref[fig:Fig2]{2(e)}, the red and blue solid lines represent the scattering intensities of the seven intrinsic CEF excitation peaks under RR and RL polarization configurations, respectively. The excitation peaks CEF1, CEF4, and CEF7 correspond to transitions from the CEF ground state to the first, fourth, and seventh excited states. These peaks exhibit significantly reduced intensities under RR polarization compared to RL polarization. This behavior aligns with Eqs. \hyperref[for:7]{(7)} and \hyperref[for:8]{(8)}, confirming that these transitions correspond to $\Gamma_4 \rightarrow \Gamma_{5,6}$ transitions. According to Eq. \hyperref[for:3]{(3)}, the $\Gamma_4 \rightarrow \Gamma_4$ transitions are governed by multiple Raman tensors, in clear contrast to the $\Gamma_4 \rightarrow \Gamma_{5,6}$ transitions. The excitation peaks CEF2, CEF3, CEF5, and CEF6 correspond to transitions from the ground state to the second, third, fifth, and sixth excited states, respectively. Except for CEF5, the polarization-dependent responses of the other three intrinsic excitations deviate significantly from the behavior expected for the $\Gamma_3$ Raman tensor. The distinct Raman polarization dependence effectively distinguishes the two symmetry-allowed transitions. The polarization behavior of CEF5 coincides with that of a $\Gamma_4 \rightarrow \Gamma_{5,6}$ transition by coincidence, which may result from the multiple Raman tensor contributions inherent to the $\Gamma_4 \rightarrow \Gamma_4$ process. Overall, the polarization-resolved Raman scattering intensities of these CEF excitations are fully consistent with the calculated symmetries of the CEF wave functions in ErOCl, providing further validation of our theoretical framework. These results highlight the crucial role of symmetry analysis in unraveling the CEF physics of rare-earth-based materials.

\begin{table*}
	\renewcommand{\arraystretch}{1.5}
	\caption{\label{tab:table1} CEF parameters of ErOCl ($\left| {J = 15/2,{J^z}} \right\rangle$).}
	\begin{ruledtabular}
		\begin{tabular}{ccccccc}
			\multicolumn{7}{c}{CEF parameters for \ce{ErOCl}}\\
			\hline
			CEF parameters & $B_{2}^{0}$ & $B_{4}^{0}$ & $B_{4}^{3}$ & $B_{6}^{0}$ & $B_{6}^{3}$ & $B_{6}^{6}$ \\
			(meV) &$9.7313 \times 10^{-3}$ & $4.4234 \times 10^{-4}$ & $1.7694 \times 10^{-2}$ & $ 6.2245 \times 10^{-6}$ & $-5.9082 \times 10^{-5}$ & $6.7576 \times 10^{-5}$ \\
			\hline
			Stevens& \( A_2^0 \langle r^2 \rangle \)  & \(  A_4^0 \langle r^4 \rangle \)  & \(  A_4^3 \langle r^4 \rangle \)  & \(  A_6^0 \langle r^6 \rangle \)  & \(  A_6^3 \langle r^6 \rangle \)  & \(  A_6^6 \langle r^6 \rangle \)  \\
			normalization & = \( B_2^0 \)/$\alpha$\textcolor{blue}{$^a$} (cm$^{-1}$)  & = \( B_4^0 \)/$\beta$\textcolor{blue}{$^b$} (cm$^{-1}$)  & = \( B_4^3 \)/$\beta$ (cm$^{-1}$)  & = \( B_6^0 \)/$\gamma$\textcolor{blue}{$^c$} (cm$^{-1}$)  & = \( B_6^3 \)/$\gamma$ (cm$^{-1}$)  & = \( B_6^6 \)/$\gamma$ (cm$^{-1}$) \\
			
			~ &$2.4727$ & $-2.0604$ & $-82.4187$ & $ 0.3392 $ & $-3.2200$ & $3.6829$  \\
			\hline
		\end{tabular}
	\end{ruledtabular}
	\label{Table:ErOClCEFInfo}
	\footnotetext[1]{$1/\alpha$ = $2.541 \times 10^2$}
	\footnotetext[2]{$1/\beta$ = $-4.658 \times 10^3$}
	\footnotetext[3]{$1/\gamma$ = $5.450 \times 10^4$}
	\label{table:TABLE I}
\end{table*}

\begin{table*}
	\centering
	\caption{ CEF energy levels, wavefunctions, and symmetry of \ce{ErOCl}}
	\renewcommand{\arraystretch}{1.5}
	\begin{tabularx}{\textwidth}{cccp{11.5cm}c} 
		\hline
		\hline
		& Exp. (meV) &\hspace{0.2cm} Fit (meV) & \hspace{4.5cm} Wavefunction &  Symmetry \\
		\hline
		\multirow{2}{*}{CEF0} & \multirow{2}{*}{0}  & \multirow{2}{*}{0} & 
		$\left| \psi_{0, \pm} \right\rangle = 
		\mp 0.0438 \left| \pm\frac{13}{2} \right\rangle 
		+ 0.5029 \left| \pm\frac{11}{2} \right\rangle 
		+ 0.0570 \left| \pm\frac{7}{2} \right\rangle 
		\mp 0.5310 \left| \pm\frac{5}{2} \right\rangle 
		\mp 0.0401 \left| \pm\frac{1}{2} \right\rangle 
		+ 0.3270 \left| \mp\frac{1}{2} \right\rangle 
		- 0.0651 \left| \mp\frac{5}{2} \right\rangle 
		\pm 0.4646 \left| \mp\frac{7}{2} \right\rangle 
		\mp 0.0617 \left| \mp\frac{11}{2} \right\rangle 
		+ 0.3570 \left| \mp\frac{13}{2} \right\rangle$ 
		&  \multirow{2}{*}{$\Gamma_4$} \\
		\multirow{2}{*}{CEF1} & \multirow{2}{*}{2.79} & \multirow{2}{*}{2.78} & $\left| \psi_{1,\pm} \right\rangle = 
		\pm 0.0208 \left| \pm\frac{15}{2} \right\rangle 
		- 0.2923 \left| \pm\frac{9}{2} \right\rangle 
		\mp 0.0392 \left| \pm\frac{3}{2} \right\rangle 
		+ 0.6939 \left| \mp\frac{3}{2} \right\rangle 
		\pm 0.6566 \left| \mp\frac{9}{2} \right\rangle 
		- 0.0004 \left| \mp\frac{15}{2} \right\rangle$ &  \multirow{2}{*}{$\Gamma_{5,6}$} \\
		\multirow{2}{*}{CEF2} & \multirow{2}{*}{3.29} & \multirow{2}{*}{3.00} & $\left| \psi_{2,\pm} \right\rangle = 
		\pm 0.2145 \left| \pm\frac{13}{2} \right\rangle 
		+ 0.2130 \left| \pm\frac{11}{2} \right\rangle 
		- 0.3453 \left| \pm\frac{7}{2} \right\rangle 
		\mp 0.0307 \left| \pm\frac{5}{2} \right\rangle 
		\pm 0.5061 \left| \pm\frac{1}{2} \right\rangle 
		- 0.1552 \left| \mp\frac{1}{2} \right\rangle 
		- 0.1001 \left| \mp\frac{5}{2} \right\rangle 
		\mp 0.1059 \left| \mp\frac{7}{2} \right\rangle 
		\mp 0.6947 \left| \mp\frac{11}{2} \right\rangle 
		- 0.0658 \left| \mp\frac{13}{2} \right\rangle$ &  \multirow{2}{*}{$\Gamma_{4}$} \\
		\multirow{2}{*}{CEF3} &  \multirow{2}{*}{4.18} & \multirow{2}{*}{4.21} & $\left| \psi_{3,\pm} \right\rangle = 
		+ 0.7696 \left| \pm\frac{13}{2} \right\rangle 
		\pm 0.0211 \left| \pm\frac{11}{2} \right\rangle 
		\pm 0.0460 \left| \pm\frac{7}{2} \right\rangle 
		- 0.1031 \left| \pm\frac{5}{2} \right\rangle 
		- 0.3900 \left| \pm\frac{1}{2} \right\rangle 
		\pm 0.2026 \left| \mp\frac{1}{2} \right\rangle 
		\mp 0.1986 \left| \mp\frac{5}{2} \right\rangle 
		+ 0.0239 \left| \mp\frac{7}{2} \right\rangle 
		- 0.0406 \left| \mp\frac{11}{2} \right\rangle 
		\mp 0.3997 \left| \mp\frac{13}{2} \right\rangle$ &  \multirow{2}{*}{$\Gamma_4$} \\
		\multirow{2}{*}{CEF4} &  \multirow{2}{*}{30.79} & \multirow{2}{*}{29.50} & $\left| \psi_{4,\pm} \right\rangle = 
		\pm 0.0010 \left| \pm\frac{15}{2} \right\rangle 
		+ 0.1099 \left| \pm\frac{9}{2} \right\rangle 
		\mp 0.1163 \left| \pm\frac{3}{2} \right\rangle 
		+ 0.4268 \left| \mp\frac{3}{2} \right\rangle 
		\mp 0.4095 \left| \mp\frac{9}{2} \right\rangle 
		- 0.7903 \left| \mp\frac{15}{2} \right\rangle$ &  \multirow{2}{*}{$\Gamma_{5,6}$} \\
		\multirow{2}{*}{CEF5} & \multirow{2}{*}{32.10} & \multirow{2}{*}{32.50} & $\left| \psi_{5,\pm} \right\rangle = 
		- 0.0084 \left| \pm\frac{13}{2} \right\rangle 
		\pm 0.4561 \left| \pm\frac{11}{2} \right\rangle 
		\mp 0.0035 \left| \pm\frac{7}{2} \right\rangle 
		+ 0.7425 \left| \pm\frac{5}{2} \right\rangle 
		- 0.0574 \left| \pm\frac{1}{2} \right\rangle 
		\pm 0.4695 \left| \mp\frac{1}{2} \right\rangle 
		\pm 0.0907 \left| \mp\frac{5}{2} \right\rangle 
		- 0.0285 \left| \mp\frac{7}{2} \right\rangle 
		- 0.0557 \left| \mp\frac{11}{2} \right\rangle 
		\pm 0.0688 \left| \mp\frac{13}{2} \right\rangle$&  \multirow{2}{*}{$\Gamma_4$} \\
		\multirow{2}{*}{CEF6} &  \multirow{2}{*}{33.17} & \multirow{2}{*}{33.42} & $\left| \psi_{6,\pm} \right\rangle = 
		\mp 0.2479 \left| \pm\frac{13}{2} \right\rangle 
		+ 0.0079 \left| \pm\frac{11}{2} \right\rangle 
		- 0.7917 \left| \pm\frac{7}{2} \right\rangle 
		\pm 0.0539 \left| \pm\frac{5}{2} \right\rangle 
		\mp 0.4339 \left| \pm\frac{1}{2} \right\rangle 
		- 0.0778 \left| \mp\frac{1}{2} \right\rangle 
		- 0.3007 \left| \mp\frac{5}{2} \right\rangle 
		\pm 0.1420 \left| \mp\frac{7}{2} \right\rangle 
		\pm 0.0442 \left| \mp\frac{11}{2} \right\rangle 
		- 0.0445 \left| \mp\frac{13}{2} \right\rangle$ &  \multirow{2}{*}{$\Gamma_4$}\\
		\multirow{2}{*}{CEF7} &  \multirow{2}{*}{34.34} & \multirow{2}{*}{34.50} & $\left| \psi_{7,\pm} \right\rangle = 
		- 0.6123 \left| \pm\frac{15}{2} \right\rangle 
		\mp 0.5379 \left| \pm\frac{9}{2} \right\rangle 
		- 0.5527 \left| \pm\frac{3}{2} \right\rangle 
		\mp 0.1260 \left| \mp\frac{3}{2} \right\rangle 
		- 0.1199 \left| \mp\frac{9}{2} \right\rangle 
		\mp 0.0002 \left| \mp\frac{15}{2} \right\rangle$ &  \multirow{2}{*}{$\Gamma_{5,6}$} \\
		\hline
		\hline
	\end{tabularx}
	\label{table:TABLE II}
\end{table*}

Building on the comprehensive analysis above, we have unequivocally determined all CEF excitations in ErOCl through Raman scattering. Importantly, Raman spectroscopy provides a unique advantage in determining the symmetry properties of these excitations and resolving high-energy CEF levels that may be difficult to detect using other techniques, such as inelastic neutron scattering. This insight lays a solid foundation for our subsequent investigation into the magnetism of ErOCl and Lu$_{x}$Er$_{1-x}$OCl across varying Lu doping concentrations, offering a deeper understanding of how chemical substitution influences CEF excitations and magnetism in this system.

\begin{figure*}[t]
	\includegraphics[scale=0.60]{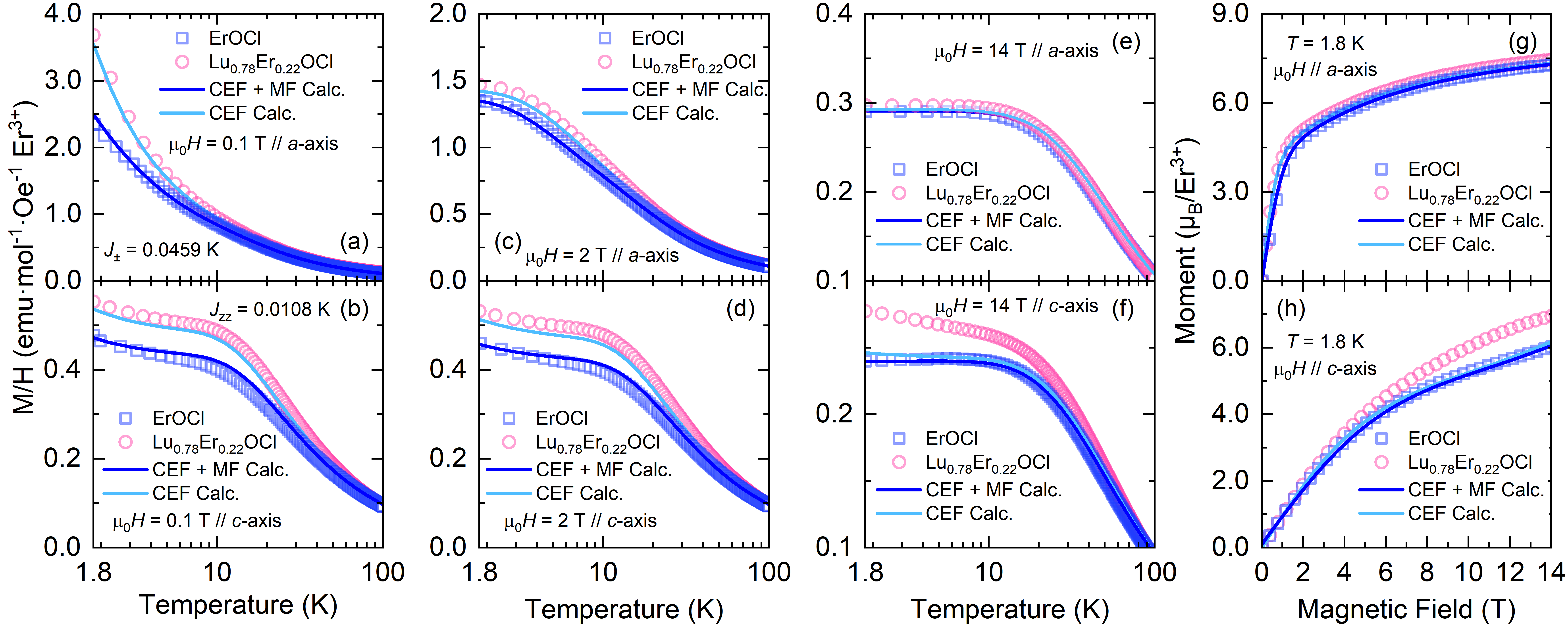}
	\caption{\textbf{Magnetization of ErOCl and Lu$_{0.78}$Er$_{0.22}$OCl.} 
		(a)-(f) Temperature dependent magnetization ($M/H$-$T$) measured under different magnetic fields along the $a$-axis and $c$-axis for ErOCl (blue open squares) and Lu$_{0.78}$Er$_{0.22}$OCl (red open circles).
		(g) and (h) Magnetic field dependent magnetization ($M$-$H$) at $T$ = 1.8 K up to 14 T along the $a$-axis and $c$-axis for ErOCl (blue open squares) and Lu$_{0.78}$Er$_{0.22}$OCl (red open circles). 
		Blue solid lines represent the MF calculation results, while the light blue solid lines represent the CEF calculation results.}
	\label{fig:Fig3}
\end{figure*}
\subsection{Magnetism of \ce{ErOCl}}
We have gained a comprehensive understanding of the CEF effects in ErOCl, enabling us to analyze its CEF-driven magnetism and directly compare it with experimentally measured magnetization data.

Fig. \hyperref[fig:Fig3]{3} presents the magnetization data of ErOCl (blue open squares) and Lu-doped Lu$_{0.78}$Er$_{0.22}$OCl (red open circles) along both the $a$-axis and $c$-axis under different magnetic fields. As shown in Figs. \hyperref[fig:Fig3]{3(a)} and \hyperref[fig:Fig3]{3(b)}, at a weak magnetic field of $\mu_0H$ = 0.1 T, the $M/H$-$T$ curves for ErOCl and Lu$_{0.78}$Er$_{0.22}$OCl exhibit deviations at low temperatures along both the $a$- and $c$-axes. As the applied magnetic field increases, the $M/H$-$T$ data along the $a$-axis for both compounds nearly coincide across the entire temperature range, as illustrated in Figs. \hyperref[fig:Fig3]{3(c)} and \hyperref[fig:Fig3]{3(e)}. In contrast, along the $c$-axis, the magnetization behavior of ErOCl and Lu$_{0.78}$Er$_{0.22}$OCl diverge more significantly at lower temperatures, as shown in Figs. \hyperref[fig:Fig3]{3(d)} and \hyperref[fig:Fig3]{3(f)}. This divergence is further emphasized in the $M$-$H$ curves at 1.8 K. As shown in Figs. \hyperref[fig:Fig3]{3(g)} and \hyperref[fig:Fig3]{3(h)}, while the magnetization behavior along the $a$-axis remains comparable, a pronounced enhancement of low temperature magnetization is observed along the $c$-axis in the Lu-doped sample. This distinct anisotropic response highlights the influence of Lu$^{3+}$ substitution on the magnetic properties of ErOCl.

To interpret the measured magnetization data, we carried out detailed theoretical calculations. Using the CEF theory and the CEF parameters listed in Table \ref{table:TABLE I}, we computed the $M/H$-$T$ curves along both the $a$- and $c$-axes under a magnetic field of 0.1 T, with the results shown as light blue solid lines in Figs. \hyperref[fig:Fig3]{3(a)} and \hyperref[fig:Fig3]{3(b)}. Interestingly, the calculated CEF magnetization data roughly align with the experimental results of the dilute magnetic sample Lu$_{0.78}$Er$_{0.22}$OCl, yet they show considerable deviations from those of ErOCl, particularly at low temperatures. This discrepancy can be understood from a physical standpoint: the substitution of Er$^{3+}$ with non-magnetic Lu$^{3+}$ increases the spacing between magnetic ions, thereby significantly weakening exchange interactions. Consequently, the magnetization behavior of Lu$_{0.78}$Er$_{0.22}$OCl is well described by the CEF model alone. By contrast, in ErOCl, spin exchange interactions become increasingly prominent at lower temperatures, making them indispensable for an accurate description of their magnetization behavior. To incorporate these effects, we employ the XXZ model with NN exchange interactions, providing a more complete theoretical framework for understanding the observed magnetization trends.
The Hamiltonian of the XXZ model is given by:
\begin{equation}
	\hat{H}_{XXZ} = \sum_{ij} \vartheta_{\pm} \left( \hat{J}_{i}^{+} \hat{J}_{j}^{-} + \hat{J}_{i}^{-} \hat{J}_{j}^{+} \right) + \vartheta_{zz} \hat{J}_{i}^{z} \hat{J}_{j}^{z},
	\label{for:9}
\end{equation}
where $\vartheta_{\pm}$ and $\vartheta_{zz}$ denote the NN exchange interactions in terms of the total angular momentum $J = 15/2$. To facilitate quantitative analysis, we applied a mean-field (MF) approximation to simplify the Hamiltonian. Within this framework, the total MF Hamiltonians along the $a$-axis and $c$-axis take the following forms \cite{52}:
\begin{equation}
	\hat{H}_{MF-a-axis} = \hat{H}_{CEF} 
	+ \frac{6 \vartheta_{\pm} M^x}{\mu_0 \mu_B g_J} \sum_i \hat{J}_i^x 
	- \mu_0 \mu_B g_J \sum_i h_x \hat{J}_i^x,
	\label{for:10}
\end{equation}
\begin{equation}
	\hat{H}_{MF-c-axis} = \hat{H}_{CEF} 
	+ \frac{3 \vartheta_{zz} M^z}{\mu_0 \mu_B g_J} \sum_i \hat{J}_i^z 
	- \mu_0 \mu_B g_J \sum_i h_z \hat{J}_i^z,
	\label{for:11}
\end{equation}
where $M^{x}$ and $M^{z}$ represent the order parameters along the $a$-axis and $c$-axis, respectively. This MF treatment effectively captures the impact of exchange interactions while maintaining computational tractability, allowing for a direct comparison with experimental magnetization data.

Building on the MF Hamiltonian outlined in Eqs. \hyperref[for:10]{(10)} and \hyperref[for:11]{(11)}, we proceeded to simulate the magnetization behavior of ErOCl. By fitting the experimental magnetization data, we obtained an optimal set of exchange interaction parameters: $\vartheta_{\pm} = 0.0459\pm0.0014 $ K and $\vartheta_{zz} = 0.0108\pm0.0006$ K. The corresponding calculated magnetization curves are presented as blue solid lines in Fig. \hyperref[fig:Fig3]{3}. Crucially, incorporating exchange interactions significantly enhanced the agreement between theoretical predictions and experimental measurements of ErOCl. This result underscored the fundamental role of exchange interactions in governing the low temperature magnetization behavior and highlighted the effectiveness of our approach in accurately describing the magnetic interactions in this system.}

The triangular-lattice quantum magnet NaYb$_{1-x}$Lu$_x$S$_2$ and NaLu$_x$Yb$_{1-x}$Se$_2$ exhibits enhanced paramagnetism with increasing Lu substitution at low temperatures and under low magnetic fields. This behavior is attributed to magnetic ion dilution, which drives the system toward divergent Curie–Weiss behavior~\cite{18,46}.
Lu$_{0.78}$Er$_{0.22}$OCl exhibits a similar paramagnetic enhancement, accompanied by a $\sim$15\% increase in the magnetic moment per Er$^{3+}$ ion. Such dilute rare-earth compounds have been widely used to calibrate CEF wavefunctions. For example, in HoB$_{12}$ and NaTmSe$_2$, the low-field magnetization of dilute magnetic samples is significantly larger than that of their undiluted counterparts, owing to the suppression of magnetization in the latter by antiferromagnetic interactions~\cite{PhysRevB.111.L180405,PhysRevB.104.134436}. Under high magnetic fields, however, these interactions are fully suppressed, and the magnetization of dilute and pure samples converges without noticeable differences. At low temperature and under high magnetic fields the measured magnetization of the dilute sample Lu$_{0.78}$Er$_{0.22}$OCl deviates markedly from the calculated magnetization based on the CEF parameters of ErOCl, as shown in Fig. \hyperref[fig:Fig3]{3(h)}. As the Curie-Weiss model does not apply under high magnetic fields, the origin of the enhanced paramagnetism in Lu$_{0.78}$Er$_{0.22}$OCl warrants further investigation.

We begin by considering Van Vleck paramagnetism as a possible origin of the enhanced paramagnetization in Lu$_{0.78}$Er$_{0.22}$OCl. Based on the preceding analysis, spin interactions play a minimal role in modulating the magnetization under high magnetic fields. Meanwhile, the magnetism of ErOCl is almost entirely governed by CEF effects [see Figs. \hyperref[fig:Fig3]{3(f)} and \hyperref[fig:Fig3]{3(h)}], indicating that Van Vleck paramagnetism is likely negligible in this compound. Given the strong similarity between ErOCl and Lu$_{0.78}$Er$_{0.22}$OCl, Van Vleck paramagnetism can be excluded as the source of the enhanced high-field paramagnetism in the doped sample.

\begin{figure*}[t]
	\includegraphics[scale=0.60]{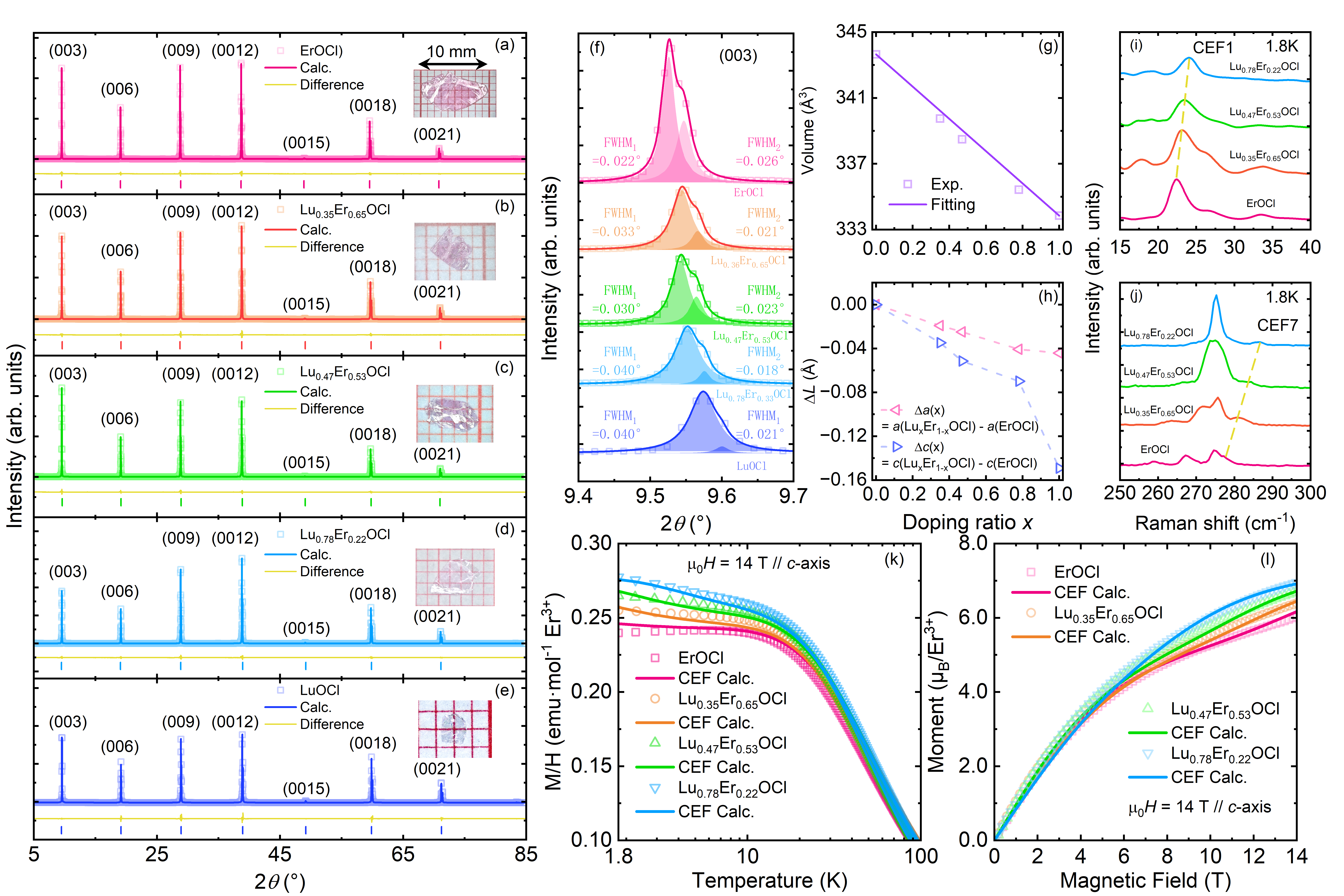}
	\caption{\textbf{Effect of Lu$^{3+}$ doping on the structural, vibrational, and magnetic properties of Lu$_x$Er$_{1-x}$OCl single crystals.} 
		(a)–(e) XRD $\left(00l\right)$ Bragg peaks, open squares represent experimental data, and solid lines are fitting results. Insets: optical images of samples with different Lu$^{3+}$ concentrations. 
		(f) Evolution of the (003) Bragg peak with increasing Lu$^{3+}$ content, showing a shift toward higher $2\theta$. 
		(g) Lattice volume as a function of Lu$^{3+}$ content; purple squares denote values from four-circle diffraction, and the solid line corresponds to Vegard’s law. 
		(h) Relative change in lattice parameters with Lu$^{3+}$ doping. 
		(i), (j) Raman spectra of the first and seventh CEF excitations at 1.8 K. 
		(k) $M/H$–$T$ and (l) $M$–$H$ curves along the $c$-axis for different Lu$^{3+}$ concentrations. Open symbols: experimental data; solid lines: calculations based on CEF theory.}
	\label{fig:Fig4}
\end{figure*}

Furthermore, low temperature Raman measurements on Lu$_x$Er$_{1-x}$OCl reveal subtle shifts in the CEF excitation peaks upon Lu substitution, as illustrated in Figs. \hyperref[fig:Fig4]{4(i)} and \hyperref[fig:Fig4]{4(j)}. A similar behavior has been reported in YbOCl~\cite{21}. Given that CEF excitations in Er-based rare-earth magnets are inherently weak, such doping-induced CEF modifications should not be overlooked. We further speculate that even minor changes in the CEF environment may have amplified effects on magnetism under high magnetic fields.
To explore how doping modulates CEF excitations and enhances high magnetic fields magnetism, we synthesized a series of Lu$_x$Er$_{1-x}$OCl samples with varying doping levels for systematic investigation.

\subsection{Low temperature magnetism enhanced by Lu$^{3+}$ doping}
A plausible explanation for the enhanced low temperature magnetism observed in the doped samples is the modification of the CEF environment compared to ErOCl. To test this hypothesis, we synthesized a series of Lu$_{x}$Er$_{1-x}$OCl single crystals with varying doping concentrations ($x$ = 0, 0.35, 0.47, 0.78, and 1). The elemental composition of these crystals was verified using X-ray energy-dispersive spectroscopy (EDS) \cite{52}, ensuring the accuracy of the doping levels. At the same time, we observed that the samples with different Lu$^{3+}$ doping concentrations exhibited only minor variations in both lattice constants and atomic coordinates. This ensures that the structural integrity of the samples remained consistent across different doping levels, thereby ruling out the possibility of structural distortions affecting the magnetic properties of the material system during the doping process.

\begin{table*}
	\renewcommand{\arraystretch}{1.5}
	\caption{\label{tab:table1} CEF parameters of Lu$_x$Er$_{1-x}$OCl in Stevens normalization. The boldface B$_2^0$ term shows the most pronounced change with $x$, reflecting the strong c-axis contraction upon doping.}
	\centering
	\begin{tabular*}{\textwidth}{@{\extracolsep{\fill}}ccccccc}
		\hline
		\hline
		$x$ & $B_{2}^{0}$ (cm$^{-1}$) & $B_{4}^{0}$ (cm$^{-1}$) & $B_{4}^{3}$ (cm$^{-1}$) & $B_{6}^{0}$ (cm$^{-1}$) & $B_{6}^{3}$ (cm$^{-1}$) & $B_{6}^{6}$ (cm$^{-1}$) \\
		\hline
		0 & \textbf{2.4727} & -2.0604 & -82.4187 & 0.3392 & -3.2200 & 3.6829 \\
		0.35 & \textbf{2.7024} & -2.0370 & -83.6996 & 0.3536 & -3.2815 & 3.6615 \\
		0.47 & \textbf{4.9626} & -2.1579 & -82.2137 & 0.3830 & -3.4787 & 3.1781 \\
		0.78 & \textbf{5.7551} & -2.0801 & -82.8006 & 0.4028 & -3.5124 & 3.0451 \\
		\hline
		\hline
	\end{tabular*}
	\label{table:TABLE III}
\end{table*}
\begin{table*}[htbp]
	\renewcommand{\arraystretch}{1.5}
	\centering
	\caption{Comparison of fitted and Raman-measured CEF of Lu$_x$Er$_{1-x}$OCl with varying $x$. }
	\begin{tabular*}{\textwidth}{@{\extracolsep{\fill}}ccccc}
		\hline
		\hline
		\( x \) & CEF1 Raman (meV) & CEF1 Fitted (meV) & CEF7 Raman (meV) & CEF7 Fitted (meV) \\
		\hline
		0.00 &  2.79 & 2.78 & 34.34 & 34.50 \\
		0.35 &  2.88 & 2.87 & 34.92 & 34.94 \\
		0.47 &  2.91 & 2.91 & 35.23 & 35.09 \\
		0.78 &  2.98 & 2.96 & 35.49 & 35.44 \\
		\hline
		\hline
	\end{tabular*}
	\label{table:TABLE IV}
\end{table*}
We conducted XRD measurements on the synthesized single-crystal samples, including four-circle diffractometer analysis. The X-ray diffraction patterns and morphological features of these single crystals are shown in Figs. \hyperref[fig:Fig4]{4(a)}-\hyperref[fig:Fig4]{4(e)}. The results reveal that the doped samples exhibit diffraction patterns nearly identical to that of ErOCl, except for systematic shifts in peak positions. This indicates that the doped samples remain isostructural with ErOCl. To further elucidate the effect of doping on lattice parameters, we performed high-resolution measurements of the $\left(003\right)$ diffraction peak, as shown in Fig. \hyperref[fig:Fig4]{4(f)}. The data clearly demonstrate a progressive shift of the $(003)$ peak toward higher $2\theta$ angles as the Lu$^{3+}$ concentration increases, transitioning from ErOCl to Lu$_{x}$Er$_{1-x}$OCl at various doping levels, and ultimately to LuOCl. This shift is attributed to the smaller ionic radius of Lu$^{3+}$ compared to Er$^{3+}$, despite their similar coordination environments. Since Lu$^{3+}$ has a higher atomic number (71 vs. 68) and a greater nuclear charge, it exerts a stronger electrostatic attraction on its outer electrons, leading to a more contracted electron cloud. Consequently, substituting Er$^{3+}$ with Lu$^{3+}$ induces lattice contraction, reducing the lattice parameters and causing the observed peak shifts in the XRD patterns toward larger $2\theta$ angles.

We further examined the evolution of the unit cell volume as a function of the Lu$^{3+}$ doping ratio based on four-circle diffraction measurements, as shown in Fig. \hyperref[fig:Fig4]{4(g)}. The data reveal a continuous shrinkage of the unit cell volume with increasing Lu$^{3+}$ concentration, following an almost linear trend that aligns well with Vegard’s law~\cite{35}:

\begin{equation}
	V(x) = (1 - x) V_{\mathrm{ErOCl}} + x V_{\mathrm{LuOCl}},
	\label{for:12}
\end{equation}

\noindent where  $x$  represents the doping ratio, $V_{\mathrm{ErOCl}}$  is the lattice volume of ErOCl, and  V$_{\mathrm{LuOCl}}$  is the lattice volume of LuOCl. The experimentally measured unit cell volumes for different doping levels closely match the theoretical predictions [purple solid line in Fig. \hyperref[fig:Fig4]{4(g)}], underscoring the high crystalline quality of the synthesized single-crystal samples. This structural consistency provides a solid foundation for exploring the impact of CEF excitations on magnetism in these materials. To gain further insight into the anisotropic structural response to Lu$^{3+}$ substitution, we compared the evolution of the lattice parameters along the $a$-axis and $c$-axis for different doping ratios, as depicted in Fig. \hyperref[fig:Fig4]{4(h)}. A key observation emerges: the contraction along the $c$-axis is significantly more pronounced than along the $a$-axis as the Lu$^{3+}$ content increases. This behavior can be attributed to the pronounced two-dimensional layered nature of ErOCl. Within this structure, the in-plane lattice is stabilized by strong ionic bonds, imparting substantial rigidity, whereas interlayer interactions are predominantly mediated by weaker van der Waals forces, rendering the $c$-axis comparatively “soft”. Consequently, the chemical pressure exerted by Lu$^{3+}$ substitution preferentially compresses this softer $c$-axis, leading to a distinctly anisotropic lattice contraction. 

This anisotropic structural response is reminiscent of similar phenomena reported in Fe-based high-temperature superconductors. For instance, in Ba$_2$Fe$_2$As$_2$-based systems~\cite{36}, substitution with alkali metals such as K or Cs introduces varying degrees of chemical pressure, which predominantly affects the $c$-axis while leaving the $a$ and $b$ axes relatively unchanged. A comparable effect has also been observed in rare-earth magnetic systems such as (Ho$_{1-x}$Lu$_{x}$)$_{2}$Fe$_{2}$Si$_{2}$C~\cite{37}, where lattice compression along the $c$-axis plays a crucial role in tuning magnetic interactions. Given that CEF parameters are highly sensitive to local lattice environment, this anisotropic structural change is expected to significantly influence the CEF excitations in ErOCl and its Lu$^{3+}$-doped variants. In particular, the pronounced contraction along the $c$-axis is likely to modify key CEF parameters, such as  $B_2^0$, which directly impacts the low temperature magnetic behavior. In the following section, we will explore in greater detail how these structural modifications translate into changes in CEF excitations and their role in enhancing magnetization at low temperatures.

Building on our structural analysis, we performed low temperature Raman scattering measurements on the doped samples to investigate how Lu$^{3+}$ substitution affects their CEF excitations, comparing the results with those of pristine ErOCl. The evolution of the first and seventh CEF excitation levels with doping concentration is presented in Figs. \hyperref[fig:Fig4]{4(i)} and \hyperref[fig:Fig4]{4(j)}, respectively. Due to doping-induced effects, the signals of other excitation levels in Raman scattering became too weak to be detected. As evident from these figures, both the first and seventh CEF excitation levels shift progressively toward higher energies as the proportion of Lu$^{3+}$ ions increases. Additionally, the scattering intensity systematically decreased with increasing Lu$^{3+}$ content, in line with our expectations. This confirms that doping effectively modified the original CEF excitations of ErOCl. Focusing on the first CEF excitation level, we observe a shift of approximately 2 wavenumbers from ErOCl to Lu$_{0.78}$Er$_{0.22}$OCl. While this absolute shift may seem modest, it is quite significant given that the first CEF excitation energy in Er-based magnets is typically weak. Even such a small change was sufficient to noticeably influence the low temperature magnetic properties of the doped samples. These results demonstrate that Lu$^{3+}$ doping induces measurable shifts in the CEF excitation levels, confirming that chemical substitution effectively modifies the local CEF environment in ErOCl. This change is sufficient to influence the low temperature magnetic properties of the doped samples, highlighting the sensitivity of CEF excitations to structural modifications.

To further elucidate the impact of Lu$^{3+}$ doping on magnetism, we conducted magnetization measurements on samples with varying doping levels and compared the results with those of ErOCl. Figs. \hyperref[fig:Fig4]{4(k)} and \hyperref[fig:Fig4]{4(l)} present the $M/H$-$T$ curves under a 14 T magnetic field and the $M$-$H$ curves at 1.8 K along the $c$-axis, respectively. By incorporating the partial CEF excitation levels obtained from Raman scattering, we analyzed the magnetization data to extract the corresponding CEF parameters for the doped samples. The CEF parameters derived from both Raman experiments and magnetization fitting for different doping ratios are summarized in Table \ref{table:TABLE III}. The magnetization curves calculated using these parameters [solid lines in Figs. \hyperref[fig:Fig4]{4(k)} and \hyperref[fig:Fig4]{4(l)}] exhibit excellent agreement with the experimental data. Furthermore, diagonalization calculations of the CEF Hamiltonian yielded energy levels that align well with those observed in Raman spectra~(See Table \ref{table:TABLE IV}). These results confirm that Lu$^{3+}$ doping effectively modifies the CEF environment, leading to an enhancement of magnetism at low temperatures.
To gain deeper insight into the effect of doping on magnetism, we examined the structural evolution at a microscopic level. The most pronounced impact of increasing Lu$^{3+}$ doping is observed along the $c$-axis, where significant lattice contraction occurs. Simultaneously, we found that the CEF parameter $B_{2}^{0}$ also exhibited a notable increase with higher doping concentrations. To understand the origin of this phenomenon, it is essential to consider the physical significance of CEF parameters. In the Stevens-operator formalism, the $B_{2}^{0} O_{2}^{0}$ term directly corresponds to the uniaxial CEF component along the principal axis of the system~\cite{7,38}—in this case, the $c$-axis in ErOCl. Consequently, $B_{2}^{0}$ is closely linked to the axial magnetic anisotropy and the degree of Ising-like behavior in the system. In contrast, higher-order CEF parameters, such as $B_{4}^{m}$ and $B_{6}^{m}$, are more sensitive to subtle distortions in the local coordination environment. Our findings suggest that in this layered material, the dominant structural response to Lu$^{3+}$ substitution is the contraction along the $c$-axis, which strongly influences these axial CEF terms. Meanwhile, the in-plane lattice parameters ($a$ and $b$) exhibit only slight contraction, preserving the symmetry of the in-plane bonding environment and resulting in relatively minor changes to CEF components associated with in-plane directions. Ultimately, magnetization measurements revealed a distinct enhancement of low temperature magnetism with increasing Lu$^{3+}$ doping, further underscoring the role of CEF modifications in tuning the magnetic properties of this system.

\subsection{Summary}
Non-magnetic doping has been widely employed as a means to tune magnetic properties, exhibiting diverse effects ranging from disrupting magnetic order to inducing novel magnetic states. 
In this work, we investigated the enhancement of magnetization per Er$^{3+}$ ion in the honeycomb-lattice antiferromagnet ErOCl upon non-magnetic doping with Lu$^{3+}$.
By systematically doping Lu$^{3+}$ into ErOCl, we observe a significant contraction of the lattice along the $c$-axis, which was particularly pronounced as the doping concentration increased.
This structural change exerted an obvious influence on the CEF excitations, notably enhancing the axial CEF parameter $B_{2}^{0}$, which governs the magnetic anisotropy along the $c$-axis.
Through the combination of XRD, Raman spectroscopy, and magnetization measurements, We demonstrate that this enhanced anisotropy leads to a remarkable increase in magnetization at low temperatures and high magnetic fields, contrary to the typical expectations of magnetic dilution. For instance, in the triangular lattice spin liquid candidate material  NaYb$_{1-x}$Lu$_x$S$_2$ and NaLu$_x$Yb$_{1-x}$Se$_2$, non-magnetic Lu$^{3+}$ doping also shows a paramagnetic enhancement at low temperatures and low magnetic fields, but this is generally attributed to the dilution effect of magnetic ions causing the system to approach Curie-Weiss behavior more closely \cite{18,46}, which differs from the anisotropy-tuning mechanism emphasized in the present study. A more direct and telling comparison comes from other rare-earth systems like HoB$_{12}$ and NaTmSe$_2$. In those materials, magnetic dilution serves to remove the suppression from antiferromagnetic interactions, thus enhancing the low-field magnetization. Crucially, however, at high magnetic fields where these interactions are overcome, the magnetization of the dilute and pure samples converges. This is in sharp contrast to our findings in ErOCl. Here, the enhancement of magnetization per Er$^{3+}$ ion persists and is even more pronounced at high fields, driven by a fundamentally different mechanism: the anisotropic c-axis contraction induced by Lu$^{3+}$ doping actively modifies the axial CEF parameter $B_{2}^{0}$ and enhances the magnetic anisotropy. These findings highlight a new approach to tune the magnetism in layered compounds by modifying CEF excitations via chemical doping, thereby establishing an effective pathway to engineer magnetic anisotropy and deepening the understanding of structure-property relationships in rare-earth magnets.

\subsection{Acknowledgments}
This work was supported by the National Key Research and Development Program of China (Grant No. 2022YFA1402704), the National Science Foundation of China (Grant No. 12274186), the Strategic Priority Research Program of the Chinese Academy of Sciences (Grant No. XDB33010100), and the Synergetic Extreme Condition User Facility (SECUF). 

\subsection{Data availability}
The data that support the findings of this study are available from the authors upon reasonable request.

\end{document}